\documentclass[%
 aip,
 amsmath,amssymb,
reprint,%
]{revtex4-1}

\usepackage{graphicx}
\usepackage{dcolumn}
\usepackage{bm}
\usepackage{xcolor}

\usepackage[utf8]{inputenc}
\usepackage[T1]{fontenc}
\usepackage{mathptmx}
\usepackage{here}

\usepackage[hang,small,bf]{caption}
\usepackage[subrefformat=parens]{subcaption}
\begin{document}


\title[Cyclotron resonance closure]{A non-local fluid closure for modeling cyclotron resonance in collisionless magnetized plasmas}

\author{Taiki Jikei}
 \email{jikei@eps.s.u-tokyo.ac.jp}
\author{Takanobu Amano}%
\affiliation{
Department of Earth and Planetary Science, The University of Tokyo\\
7-3-1 Hongo, Bunkyo-ku, Tokyo, 113-0033, Japan}

\date{\today}

\begin{abstract}
A fluid description for collisionless magnetized plasmas that takes into account the effect of cyclotron resonance has been developed. Following the same approach as the Landau fluid closure, the heat flux components associated with transverse electromagnetic fluctuations are approximated by a linear combination of lower-order moments in wavenumber space. The closure successfully reproduces the linear cyclotron resonance for electromagnetic waves propagating parallel to the ambient magnetic field. In the presence of finite temperature anisotropy, the model gives approximately correct prediction for an instability destabilized via the cyclotron resonance. A nonlinear simulation demonstrates the wave growth consistent with the linear theory followed by the reduction of initial anisotropy, and finally, the saturation of the instability. The isotropization may be understood in terms of quasi-linear theory, which is developed within the framework of the fluid model but very similar to its fully kinetic counterpart. The result indicates that both linear and nonlinear collisionless plasma responses are approximately incorporated in the fluid model.
\end{abstract}

\maketitle

\section{Introduction}
\label{sec1}
Physical models of collisionless plasmas can be roughly divided into two types: fluid and kinetic models. The magneto-hydrodynamics (MHD) model is by far the most widely-used model in the plasma community. It describes large-scale and low-frequency plasma phenomena and is suitable for macroscopic numerical simulations. The drawback of MHD is that it completely ignores the small-scale kinetic physics, which sometimes plays a role even for the global dynamics. One such example is magnetic reconnection. Although it is essentially an MHD phenomenon, the dynamics of collisionless magnetic reconnection may be governed by small-scale physics that breaks the frozen-in condition in the close vicinity of the diffusion region. Therefore, even the global consequence might be regulated largely by kinetic physics. Such kinetic effect at a small scale is often modeled phenomenologically in MHD by using \textit{ad hoc} anomalous resistivity. The validity of such treatment has always been a matter of debate. It is possible to extend MHD to include the Hall and the finite electron inertial effects within the fluid model framework. However, a fluid model in the conventional sense lacks collisionless wave-particle interactions in hot plasmas such as Landau and cyclotron resonances.

Kinetic models, on the other hand, solve the Vlasov equation at least for one of the species and can simulate fully collisionless dynamics. The particle-in-cell (PIC) method is the most popular simulation method, which employs a large number of super-particles to represent the velocity distribution function. PIC naturally requires much more computational power than fluid models because of the large degrees of freedom to represent arbitrary distribution functions. Furthermore, a fully kinetic explicit PIC code, in which both ions and electrons are treated as particles, has a severe limitation for the grid size. The Vlasov simulation, which is free from the limitation on the grid size, solves the Vlasov equation directly using mesh in phase space and is an alternative to the PIC method. Although it has a better convergence property than a PIC code that includes discrete particle noise, the additional mesh in velocity space increases the required computational resources even further. In any case, a fully kinetic treatment demands prohibitively large computational power for accurate and long-term numerical modeling for macroscopic phenomena.

The problem in collisionless plasma modeling is that one has to adopt either a fluid or fully kinetic model in the first place. Once the choice has been made, conventional fluid models ignore much of the collisionless effect, whereas the fully kinetic model tries to include everything. The huge gap in the degree of approximation between them makes it difficult to interpret the differences. It would be desirable to develop an intermediate model in which the kinetic effects such as wave-particle interactions are included approximately, possibly with a modest increase in the computational cost, so that macroscopic numerical modeling is still possible. In addition, one might hope that such a model ease the complexity of theoretical analyses because of reduced degrees of freedom.

One of the possible directions is to incorporate some of the kinetic effects approximately into a fluid model. \citet{Hammett1990} developed a scheme to take into account the collisionless Landau damping effect for electrostatic waves in a fluid model. It is well known that a fluid model can be systematically derived from the Vlasov equation by taking velocity moments, which yields an infinite series of moment hierarchy. The moment equations must be closed by appropriate choice of closure relations. The resonant wave-particle interaction effect will be completely lost unless the closure is carefully designed \citep{Hammett1992}. Physically, the resonance effect is associated with collisionless phase mixing (or particle free streaming) that requires a non-local closure. The so-called Landau closure proposed by \citet{Hammett1990} approximates the highest moment in the hierarchy (as a result of truncation) by a linear combination of lower-order moments in wavenumber space so that it retains non-local nature. By choosing appropriate coefficients such that the fluid response against perturbation approximates the fully kinetic response, one may obtain a linear dispersion relation that includes the Landau damping effect.

The Landau closure has also been incorporated into anisotropic MHD in the low-frequency limit \citep{Snyder1997}. Such a model sometimes referred to as kinetic MHD or collisionless MHD, better describes low-frequency linear kinetic instabilities. For instance, the non-local heat flux parallel to the ambient magnetic field as represented by the Landau closure algorithm has been shown to yield the correct threshold for the mirror instability, which is not the case in the standard Chew-Goldberger-Low (CGL) equation of state (EoS) \citep{Chew1956}. The realistic modeling capability of the temperature anisotropy effect is believed to be important for the collisionless system even at macroscopic scale sizes. One of the possible astrophysical applications is the dynamics of collisionless accretion disks around black holes, which has been investigated using a collisionless MHD model \citep{Sharma2003,Sharma2006}.

Although various extensions of the Landau closure beyond MHD have been proposed \citep{Goswami2005,Passot2007,Passot2012,Sulem2015}, none of them are able to reproduce the cyclotron resonance. It is perhaps because many of these models are mainly intended for application to fusion plasmas, where the background magnetic field is so strong and high-frequency electromagnetic fluctuations are not relevant. On the other hand, space and astrophysical plasmas are often in a high plasma beta state, and the cyclotron resonance appears to be of great importance in many applications. In this paper, we present a model to take into account the cyclotron resonance effect for electromagnetic waves propagating parallel to the ambient magnetic field in a collisionless plasma. We have applied the same procedure as the original Landau closure to the heat flux components associated with the transverse electromagnetic fluctuations. We demonstrate that the resulting dispersion relation approximately reproduces the cyclotron damping effect. The model is also able to predict an instability driven by finite temperature anisotropy. Furthermore, nonlinear simulation of the anisotropy-driven instability shows that the model describes not only the linear growth phase but also the relaxation and the saturation in the nonlinear phase. The relaxation process may be understood in terms of quasi-linear theory. We argue that incorporating a non-local closure is essential to describe linear and nonlinear collisionless plasma responses.

This paper is organized as follows. In section \ref{sec2}, we derive a fluid model by taking velocity moments of the Vlasov equation and point out the importance of the closure for the fluid-Maxwell coupled system of equations. The non-local closure model is introduced in section \ref{sec3}. Nonlinear simulation results obtained with the proposed model are discussed in section \ref{sec4}. Finally, discussion and conclusions are given in section \ref{sec5}.

\section{Fluid models of plasmas}
\label{sec2}
We start from the Vlasov equation describing the evolution of the velocity distribution function $f_s = f_s(\bm{x}, \bm{v})$ for a particle species $s$ ($i$ for ions and $e$ for electrons) with charge $e_s$ and mass $m_s$:
\begin{equation}
\frac{\partial f_s}{\partial t}+\bm{v}\cdot\nabla f_s+\frac{e_s}{m_s}(\bm{E}+\bm{v}\times\bm{B})\frac{\partial f_s}{\partial \bm{v}}=0,
\label{Vlasoveq}
\end{equation}
where $\bm{E}$ and $\bm{B}$ are the electric and magnetic fields, respectively. We define the following moment quantities: the number density $n_s=\int f_s d^3v$, bulk velocity $\bm{u}_s=\int \bm{v} f_s d^3v/n_s$, pressure tensor $\bm{p}_s=m_s \int (\bm{v}-\bm{u}_s)(\bm{v}-\bm{u}_s) f_s d^3v$, and heat flux tensor $\bm{q}_s=m_s \int (\bm{v}-\bm{u}_s)(\bm{v}-\bm{u}_s)(\bm{v}-\bm{u}_s) f_s d^3v$. Taking velocity moments of the Vlasov equation up to second order, we obtain the following equations:
\begin{eqnarray}
\frac{\partial n_s}{\partial t}+\nabla\cdot(n_s\bm{u}_s)=0,
\label{continuity}\\
\frac{\partial \bm{u}_s}{\partial t}+\bm{u_s}\cdot\nabla\bm{u_s}
+\frac{1}{m_sn_s}\nabla\cdot\bm{p}_s-\frac{e_s}{m_s}(\bm{E}+\bm{u}_s\times\bm{B})=0,
\label{EOM} \\
\frac{\partial \bm{p}_s}{\partial t}+\nabla\cdot(\bm{p}_s\bm{u}_s+\bm{q}_s)
+\left(\bm{p}_s\cdot\nabla\bm{u}_s+\frac{e_s}{m_s}\bm{B}\times\bm{p}_s\right)^{\mathcal{S}}=0,
\label{pressureeq}
\end{eqnarray}
where the superscript $^{\mathcal{S}}$ denotes symmetrization of a tensor: $\bm{a}^{\mathcal{S}}=\bm{a}+\bm{a}^{\mathsf{T}}$. Note that we define the divergence of a tensor by $(\nabla\cdot\bm{A})=\partial_N A_{ijk\ldots N}$, i.e., the summation is taken along the last index.

Maxwell's equations are given by
\begin{eqnarray}
\nabla\cdot\bm{B}=0, \label{M1} \\
\frac{\partial \bm{B}}{\partial t}=-\nabla\times\bm{E}, \label{M2}\\
\nabla\cdot\bm{E}=\frac{\rho}{\epsilon_0}, \label{M3} \\
\nabla\times\bm{B}=\mu_0\left(\bm{j}+\epsilon_0\frac{\partial \bm{E}}{\partial t}\right), \label{M4}
\end{eqnarray}
where $\epsilon_0$ and $\mu_0$ are the permittivity and permeability in vacuum, which define the speed of light $c=1/\sqrt{\epsilon_0 \mu_0}$. The fluid quantities and the electromagnetic fields are coupled through the charge and current densities defined by
\begin{eqnarray}
\rho=\sum_s e_sn_s, \\
\bm{j}=\sum_s e_sn_s\bm{u}_s.
\end{eqnarray}
Once the charge and current densities are given, Maxwell's equations can readily be solved.

The moment equations (\ref{continuity}-\ref{pressureeq}), however, have a number of unknowns larger than the number of equations. In other words, there is no equation to determine the heat flux $\bm{q}$. Although one could proceed to take the third-order moment of the Vlasov equation to obtain the equation for the heat flux, the fourth-order moment quantity will appear in the equation. In this way, the moment hierarchy will never close by itself, which is called the closure problem. One has to truncate the hierarchy at some point and adopt some assumptions to determine the highest-order moment. It is important to understand that the infinite series of moment equations and the original Vlasov equation are mathematically equivalent. It is the closure assumption involving truncation of the hierarchy that introduces the crucial difference between the moment equations and the fully kinetic equation. Therefore, developing a good closure model is a reasonable and straightforward strategy to incorporate the kinetic effects into a fluid model.

The simplest possible choice is to ignore the heat flux and assume an EoS to determine the pressure. The adiabatic EoS (or the polytropic law with a polytropic index $\gamma$) has been the most widely-used one to determine the scalar pressure:
\begin{equation}
\frac{d}{dt}\left(\frac{p_s}{n_s^{\gamma}} \right)=0,
\end{equation}
where the scalar pressure is defined by $p_s = \mathrm{Tr}(\bm{p}_s)/3$. Most of MHD, Hall-MHD, two-fluid, as well as multi-fluid codes routinely used today in numerical plasma modeling, are based on the simplest EoS. Note that the closure problem is quite generic and inherent in the fluid model, which can be applied independently to any species one-by-one. It is nothing to do with other approximations such as the charge neutrality assumption, various approximations for the generalized Ohm's law, etc.

One may relax the assumption of the scalar pressure. For instance, if the pressure is gyrotropic (or symmetric in the plane perpendicular to the magnetic field), the parallel and perpendicular (with respect to the magnetic field) pressures may evolve independently. The CGL EoS, which assumes the conservation of first and second adiabatic invariants, is a popular choice to determine the anisotropic pressure. It is even possible to consider the evolution of finite non-gyrotropic components, which are known to play a role in collisionless magnetic reconnection \citep{Hesse1993,Hesse1994}.

We should emphasize that none of these models mentioned above take into account the collisionless wave-particle interaction effect which requires a non-local closure. The Landau closure takes into account the non-local nature of collisionless plasmas and is able to approximate the Landau resonance effect in a fluid model. However, existing models derive the closure relation considering only the particle motion parallel to the magnetic field in the electrostatic limit or in electromagnetic fluctuations in the low-frequency limit using the guiding-center approximation. Consequently, there has been no model that includes the cyclotron resonance effect, which obviously involves both high-frequency electromagnetic fluctuations and fast cyclotron motion.

In the following section, we use the moment equations (\ref{continuity}-\ref{pressureeq}) involving the full $3\times3$ pressure tensor and the underlying full Vlasov equation (\ref{Vlasoveq}) for constructing the closure. In contrast to previous works, no approximation has been made on the wave frequency or the particle trajectory in the basic equations. We see that the same approach as the Landau closure applied to the basic equations yields a reasonable approximation to the cyclotron resonance effect as naturally expected.

\section{Closure in wavenumber space}
\label{sec3}

We now discuss the closure to determine the heat flux $\bm{q}$. Note that the subscript $s$ for particle species will be omitted whenever obvious. We let the ambient magnetic field be in the $z$ direction $\bm{B}_0=B_0\bm{e}_z$ and consider parallel propagation $\bm{k}=k\bm{e}_z$ which is equivalent to $\partial/\partial x=\partial/\partial y=0$. We work in the rest frame of the fluid $(\bm{u}_0=\bm{0})$ and assume that the plasma is homogeneous in the unperturbed state with a gyrotropic pressure: $p_{xy} = p_{xz} = p_{yz} = 0, p_{xx} = p_{yy} = p_{\perp0}, p_{zz} = p_{\parallel0}$. The linearized set of moment equations (\ref{continuity}-\ref{pressureeq}) are then given by:
\begin{align}
 & \frac{\partial \tilde{n}}{\partial t} +
 n_0\frac{\partial \tilde{u}_z}{\partial z}=0,
 \label{cont}\\
 & \frac{\partial \tilde{u}_x}{\partial t} +
 \frac{1}{mn_0}\frac{\partial \tilde{p}_{xz}}{\partial z} -
 \frac{e}{m}(\tilde{E}_x+\tilde{u}_yB_0)=0,
 \label{eomx}\\
 & \frac{\partial \tilde{u}_y}{\partial t} +
 \frac{1}{mn_0}\frac{\partial \tilde{p}_{yz}}{\partial z} -
 \frac{e}{m}(\tilde{E}_y-\tilde{u}_xB_0)=0,
 \label{eomy}\\
 & \frac{\partial \tilde{u}_z}{\partial t} +
 \frac{1}{mn_0}\frac{\partial \tilde{p}_{zz}}{\partial z} -
 \frac{e}{m}\tilde{E}_z=0,
 \label{eomz}\\
 & \frac{\partial}{\partial t} \tilde{p}_{xx}
 + p_{\perp0} \frac{\partial}{\partial z} \tilde{u}_{z}
 + \frac{\partial}{\partial z} \tilde{q}_{xxz}
 - 2 \frac{e}{m} B_0 \tilde{p}_{yx} = 0,
 \label{pxx}\\
 & \frac{\partial}{\partial t} \tilde{p}_{yy}
 + p_{\perp0} \frac{\partial}{\partial z} \tilde{u}_{z}
 + \frac{\partial}{\partial z} \tilde{q}_{yyz}
 + 2 \frac{e}{m} B_0 \tilde{p}_{xy} = 0,
 \label{pyy}\\
 & \frac{\partial}{\partial t} \tilde{p}_{zz}
 + 3 p_{\parallel0} \frac{\partial}{\partial z} \tilde{u}_{z}
 + \frac{\partial}{\partial z} \tilde{q}_{zzz} = 0,
 \label{pzz}\\
 & \frac{\partial}{\partial t} \tilde{p}_{xy}
 + \frac{\partial}{\partial z} \tilde{q}_{xyz}
 + \frac{e}{m} B_0 \left( \tilde{p}_{xx} - \tilde{p}_{yy} \right) = 0,
 \label{pxy}\\
 & \frac{\partial}{\partial t} \tilde{p}_{xz}
 + p_{\parallel0} \frac{\partial}{\partial z} \tilde{u}_{x}
 + \frac{\partial}{\partial z} \tilde{q}_{xzz}
 - \frac{e}{m} \tilde{B}_{y} \left( p_{\perp0} - p_{\parallel0} \right)
 - \frac{e}{m} B_0 \tilde{p}_{yz} = 0,
 \label{pxz}\\
 & \frac{\partial}{\partial t} \tilde{p}_{yz}
 + p_{\parallel0} \frac{\partial}{\partial z} \tilde{u}_{y}
 + \frac{\partial}{\partial z} \tilde{q}_{yzz}
 + \frac{e}{m} \tilde{B}_{x} \left( p_{\perp0} - p_{\parallel0} \right)
 + \frac{e}{m} B_0 \tilde{p}_{xz} = 0.
 \label{pyz}
\end{align}
All the first-order perturbations are denoted by $~\tilde{}~$. Observe that these equations may be split into the longitudinal and transverse components: The longitudinal components involve the field-aliened flow $\tilde{u}_z$ whereas the transverse components are independent of that.

The dynamics associated with electrostatic waves are described by $\tilde{n}, \tilde{u}_z, \tilde{p}_{zz}$ for which the original Landau fluid closure proposed by \citet{Hammett1990} may be used. The perpendicular pressure perturbations $\tilde{p}_{xx}, \tilde{p}_{yy}, \tilde{p}_{xy}$ (recall the symmetry of pressure tensor $\tilde{p}_{xy} = \tilde{p}_{yx}$) are dependent only passively on $\tilde{u}_z$. They describe essentially the cyclotron motion of the perturbed pressure, which decouple from the other dynamics unless physically unreasonable assumptions on the heat flux $\tilde{q}_{xxz}, \tilde{q}_{yyz}, \tilde{q}_{xyz}$ are adopted. In the following, we assume that the transverse components $\tilde{u}_{x}, \tilde{u}_{y}, \tilde{p}_{xz}, \tilde{p}_{yz}$ are decoupled from the longitudinal dynamics.

At first, we consider the case with an isotropic zeroth-order pressure tensor $\bm{p}_{0}=p_{0}\bm{1}$; extension to the anisotropic case $(p_{\perp0}\neq p_{\parallel0})$ will be discussed later. Fourier transformation of Eqs. (\ref{eomx}), (\ref{eomy}), (\ref{pxz}), and (\ref{pyz}) gives

\begin{eqnarray}
-i\omega \tilde{u}_x + ik\tilde{p}_{xz}/mn_0-\frac{e}{m}(\tilde{E}_x+\tilde{u}_yB_0)=0,\\
-i\omega \tilde{u}_y + ik\tilde{p}_{yz}/mn_0-\frac{e}{m}(\tilde{E}_y-\tilde{u}_xB_0)=0,\\
-i\omega \tilde{p}_{xz}+ik(p_0\tilde{u}_x+\tilde{q}_{xzz})-\frac{eB_0}{m}\tilde{p}_{yz}=0,\\
-i\omega \tilde{p}_{yz}+ik(p_0\tilde{u}_y+\tilde{q}_{yzz})+\frac{eB_0}{m}\tilde{p}_{xz}=0.
\end{eqnarray}
We assume the following form of $\tilde{q}_{xzz}$ and $\tilde{q}_{yzz}$:
\begin{equation}
\begin{cases}
\tilde{q}_{xzz}=\Pi\tilde{u}_x+\nu\tilde{p}_{xz},\\
\tilde{q}_{yzz}=\Pi\tilde{u}_y+\nu\tilde{p}_{yz},
\end{cases}
\label{defhf}
\end{equation}
here $\Pi$ and $\nu$ are free parameters having dimensions of pressure and velocity, respectively.
Although this form is not unique,
this results in decomposition of eigenmodes corresponding to the cold plasma limit after diagonalization.
At this point, they can be functions of wavenumber $k$ but we later find that only the sign of $\nu$ changes depending on the sign of $k$. From these equations, we obtain the relationship between $(\tilde{u}_x\,\tilde{u}_y)^{\mathsf{T}}$ and $(\tilde{E}_x\,\tilde{E}_y)^{\mathsf{T}}$:
\begin{equation}
\begin{pmatrix}
-i\omega' & -\Omega'\\
\Omega' & -i\omega'
\end{pmatrix}
\begin{pmatrix}
\tilde{u}_x\\
\tilde{u}_y
\end{pmatrix}
=
\frac{e}{m}
\begin{pmatrix}
\tilde{E}_x\\
\tilde{E}_y
\end{pmatrix},
\end{equation}
where we have introduced the definition
\begin{equation}
\begin{cases}
\omega'=\omega\left(1-\frac{p_{0}+\Pi}{n_0m}\frac{\omega-k\nu}{\omega}\frac{k^2}{(\omega-k\nu)^2-\Omega^2}\right),\\
\Omega'=\Omega\left(1+\frac{p_{0}+\Pi}{n_0m_s}\frac{k^2}{(\omega-k\nu)^2-\Omega^2}\right),
\end{cases}
\end{equation}
with $\Omega=eB_0/m$ being the cyclotron frequency. Notice that the cold plasma result is recovered in the
cold or long-wavelength limit ($p_0+\Pi\to 0$ or $k\to 0$). We may obtain the current density by calculating the inverse matrix and taking the sum over all species. We introduce quantities such as
$\hat{q}=\tilde{q}_{xzz}+i\tilde{q}_{yzz}$, $\hat{u}=\tilde{u}_x+i\tilde{u}_y$,
and $\hat{p}=\tilde{p}_{xz}+i\tilde{p}_{yz}$
to diagonalize the matrix, which, together with Maxwell's Eqs. (\ref{M2}), (\ref{M4}), gives the following dispersion relation.
\begin{equation}
1-\frac{c^2k^2}{\omega^2}-\sum_s \frac{\omega_{ps}^{2}}{\omega(\omega'_s-\Omega'_s)}=0,
\end{equation}
where $\omega_{ps}=\sqrt{(n_0e_s^2)/(m_s\epsilon_0)}$ is the plasma frequency. In this form, positive (negative) $\mathrm{Re}[\omega]$ corresponds to left-hand (right-hand) circular polarization, respectively. Note that Eq. (\ref{defhf}) can be rewritten as follows
\begin{equation}
\hat{q}=\Pi\hat{u}+\nu\hat{p}. \label{redefhf}
\end{equation}

The corresponding fully kinetic dispersion relation is given by
\begin{equation}
1-\frac{c^2k^2}{\omega^2}+\sum_s \frac{\omega_{ps}^{2}}{\sqrt{2}\omega|k|v_{\mathrm{th},s}}
Z\left(\frac{\omega-\Omega_s}{\sqrt{2}|k|v_{\mathrm{th},s}}\right)=0,
\end{equation}
where $v_{\mathrm{th},s}=\sqrt{p_{0,s}/(n_0m_s)}$ is the thermal velocity and $Z(\zeta)$ is the plasma dispersion function with argument $\zeta$ \cite{Stix}. Therefore, our task is to find the values of $\Pi$ and $\nu$ so that
\begin{equation}
-\frac{1}{(\omega'_s-\Omega'_s)}
\sim\frac{1}{\sqrt{2}|k|v_{\mathrm{th},s}}Z\left(\frac{\omega-\Omega_s}{\sqrt{2}|k|v_{\mathrm{th},s}}\right)
\end{equation}
is approximately satisfied.
Writing $a=k\nu_s/(\sqrt{2}|k|v_{\mathrm{th},s}),b=1/2+\Pi_s/(2p_{0,s})$ and $\zeta_s=(\omega-\Omega)/(\sqrt{2}|k|v_{\mathrm{th},s})$, we see
\begin{equation}
-\frac{1}{(\omega'_s-\Omega'_s)}=-\frac{1}{\sqrt{2}|k|v_{\mathrm{th},s}}\frac{\zeta_s-a}{\zeta_s^2-a\zeta_s-b}.
\end{equation}
By matching the first two terms of the Taylor series expansion of the rational function and $Z(\zeta)$
\begin{equation}
Z(\zeta)=i\sqrt{\pi}-2\zeta-\cdots ,
\end{equation}
we obtain
\begin{equation}
\begin{cases}
a=-\frac{i\sqrt{\pi}}{\pi-2},\\
b=\frac{1}{\pi-2}.
\end{cases}
\end{equation}
For $|\zeta|\gg1$, we have already seen that this model approaches the cold plasma limit which is consistent with the asymptotic form of $Z(\zeta)$
\begin{equation}
Z(\zeta)=-\frac{1}{\zeta}-\frac{1}{2\zeta^3}-\cdots .
\end{equation}

\begin{figure}[H]
\begin{center}
\includegraphics[width=0.75\linewidth]{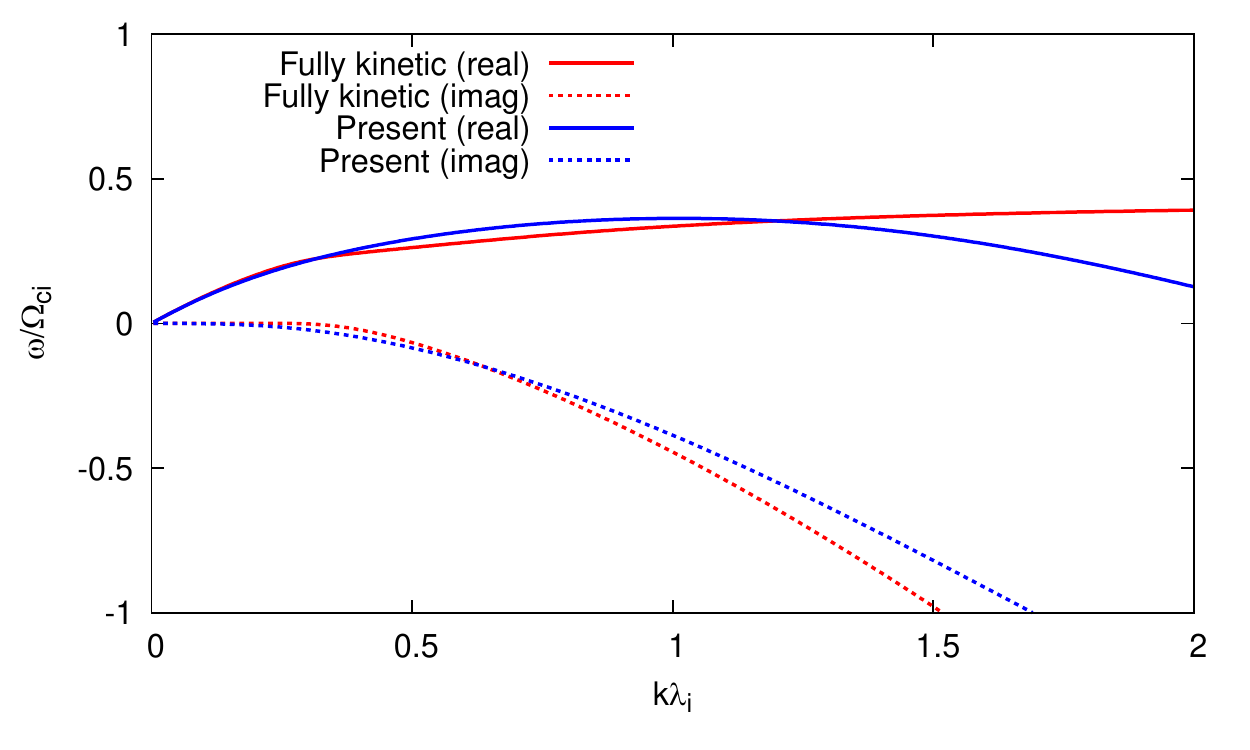}
\caption{\label{l} Dispersion relation of left-handed polarized mode obtained with $\beta_i=\beta_e=1$.
The blue color indicates the result obtained by the proposed closure method, whereas the red color indicates the fully kinetic theory.}
\end{center}
\end{figure}

\begin{figure}[H]
\begin{center}
\includegraphics[width=0.75\linewidth]{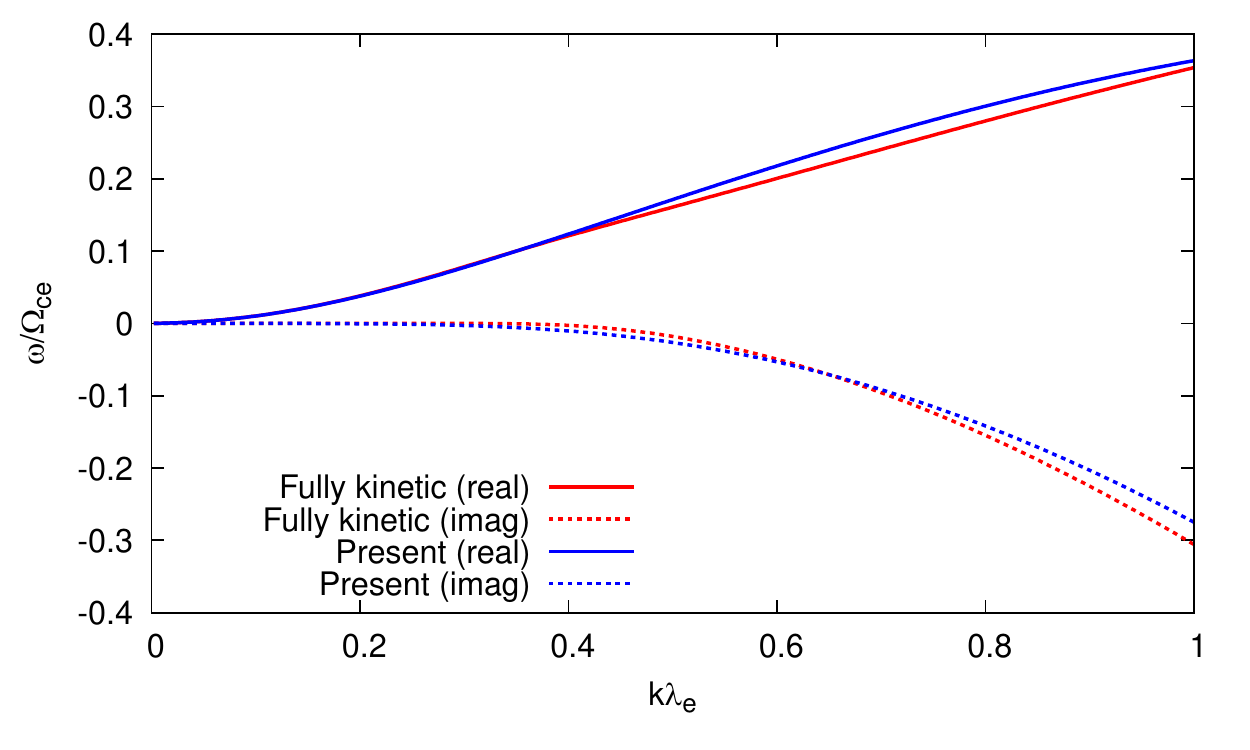}
\caption{\label{r} Dispersion relation of right-handed polarized mode obtained with $\beta_i=\beta_e=1$ with the format same as FIG. \ref{l}.
Note that negative frequency $-\mathrm{Re}[\omega]$ is plotted for the real part.}
\end{center}
\end{figure}

The resulting dispersion relation for $m_i/m_e=1836, \beta_i=\beta_e=1, (\beta_s=p_{0,s}/(B_0^2/2\mu_0))$ is shown and compared with the fully kinetic results in FIG. \ref{l},\ref{r}. Inertial length of each species are denoted by $\lambda_s=c/\omega_{ps}$. Note that we have assumed $kc/\omega\gg1$ which is a good approximation if $\omega_{ps}/\Omega_s\gg1$ is satisfied for both ions and electrons.  They both capture the effect of cyclotron damping characteristics of collisionless magnetized plasmas. Notice the normalization and scales of both axes are different in these figures.
\begin{figure}[H]
\begin{center}
\includegraphics[width=0.75\linewidth]{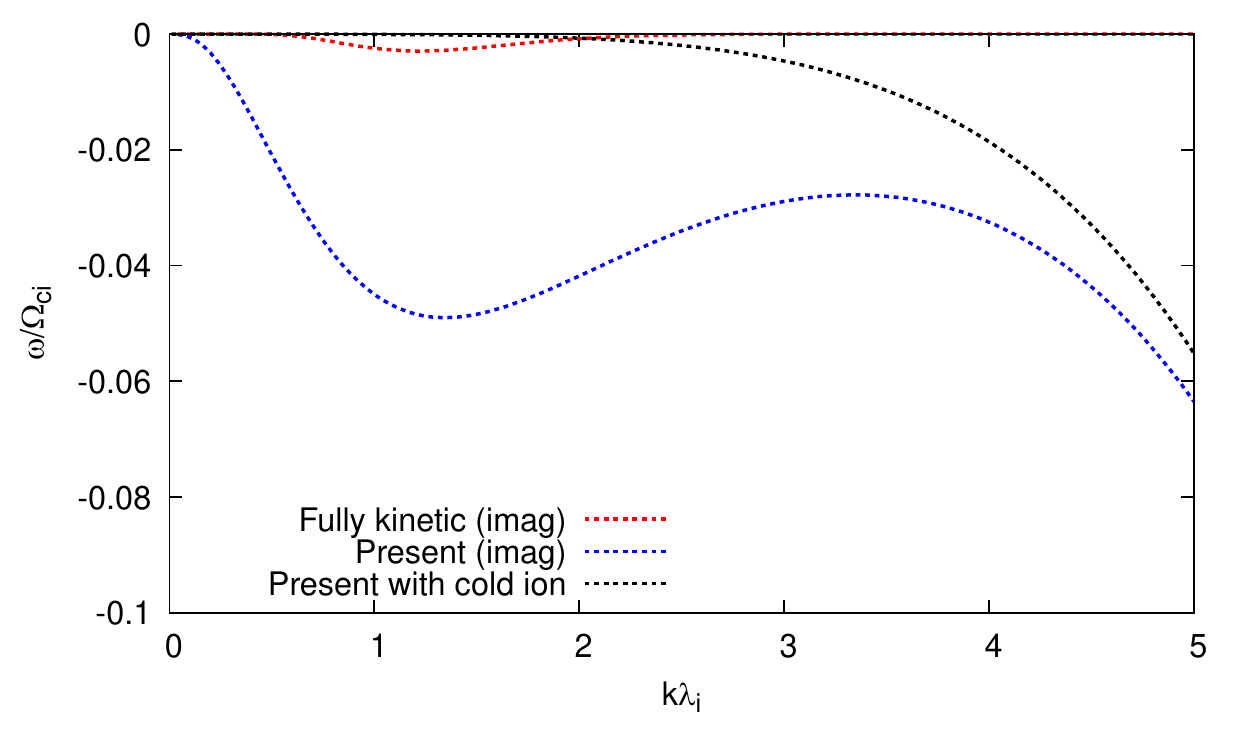}
\caption{\label{anomalous} Damping rate of right-handed polarized mode near ion inertial wavelength obtained with $\beta_i=\beta_e=1$.
The black dashed line indicates the result obtained with the cold ions $(\beta_i=0)$.}
\end{center}
\end{figure}
Let us see the details of the damping rate for the right-handed polarized mode. FIG. \ref{anomalous} provides the enlarged view for the damping rate of the right-handed polarized mode at wavelengths comparable to the ion inertial length. The damping rate obtained with the cold ion approximation $(\beta_i=0)$ is also shown in black for comparison. From this, we see the damping at this scale is associated with the ion cyclotron resonance. Although the closure model gives a somewhat larger damping rate, it correctly predicts the qualitative characteristics of the mode. It is rather more important that the error in this model keeps the system stable because otherwise, such a model can be unstable for nonlinear simulations where multiple different modes coexist and interact with each other. In section \ref{sec4}, we will see that a nonlinear simulation based on this model has been conducted without any unwanted instabilities.

\begin{figure}[H]
\begin{center}
\includegraphics[width=0.75\linewidth]{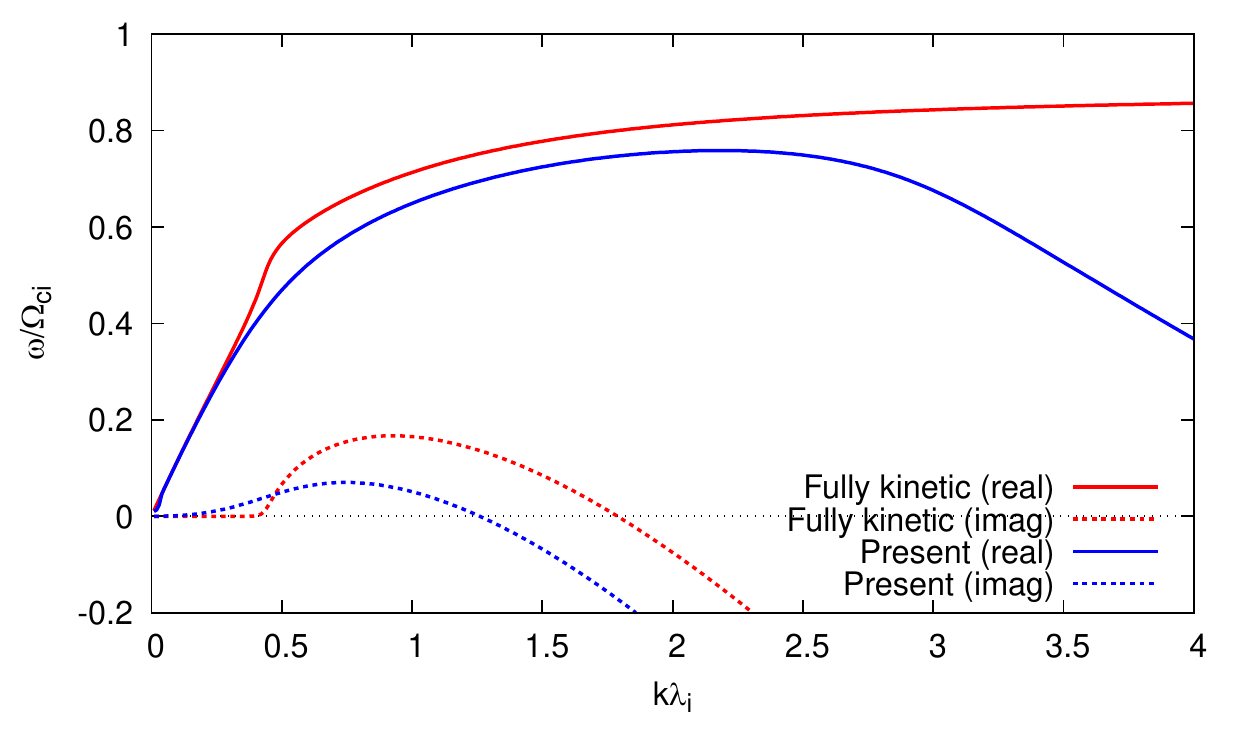}%
\caption{\label{emic} Dispersion relation of EMIC anisotropy instability obtained with $\beta_{\perp,i}=1,\beta_{\parallel,i}=0.2$.
The format is the same as FIG. \ref{l}}
\end{center}
\end{figure}

Let us now consider finite zeroth-order pressure anisotropy:
\begin{equation}
\bm{p}_{0,s}=\mathrm{diag}(p_{\perp0,s},p_{\perp0,s},p_{\parallel0,s}).
\end{equation}
Note that, as we can see from Eqs.~(\ref{pxz}-\ref{pyz}), this introduces extra terms into the equations for $\tilde{p}_{xz}, \tilde{p}_{yz}$. With the same closure ($p_{\parallel0}$ now replaces $p_0$), we have the following dispersion relation:
\begin{equation}
1-\frac{c^2k^2}{\omega^2}-\sum_s \frac{\omega_{ps}^{2}}{\omega}
\left(\frac{1}{\sqrt{2}|k|v_{\mathrm{th},s}}\frac{\zeta_s-a}{\zeta_s^2-a\zeta_s-b}
+\frac{A_s}{2\omega}\frac{1}{\zeta_s^2-a\zeta_s-b}\right)=0,\label{anisodisp}
\end{equation}
where $A_s=p_{\perp 0,s}/p_{\parallel 0,s}-1$ represent the anisotropy.
The fully kinetic dispersion relation for parallel propagation in bi-Maxwellian plasmas is given by \cite{Stix}
\begin{equation}
1-\frac{c^2k^2}{\omega^2}+\sum_s \frac{\omega_{ps}^{2}}{\omega}
\left[\frac{1}{\sqrt{2}|k|v_{\mathrm{th},s}}Z(\zeta_s)
+\frac{A_s}{\omega}\left(1+\zeta_s Z(\zeta_s)\right)\right]=0.\label{kinemic}
\end{equation}
Comparison between Eq. (\ref{anisodisp}) and (\ref{kinemic}) is shown in FIG. \ref{emic} with the same parameters used by \citet{Davidson1975}: $m_i/m_e=1836, \beta_{\perp,i}=1, \beta_{\parallel,i}=0.2, 3\beta_e/2=\beta_{\perp,i}+\beta_{\parallel,i}/2$ (the result is not sensitive to electron parameters).
Although the Taylor expansion of the second term introduced by finite anisotropy $(1/(\zeta_s^2-a\zeta_s-b))$ does not agree with the corresponding term obtained from the fully kinetic theory $(1+\zeta_s Z(\zeta_s))$ even for the lowest order term, the closure model gives a qualitatively correct result. In principle, it is possible to modify the closure for a better approximation for finite $A\neq0$. However, that would introduce extra error on the first term, which we find undesirable. As we will see in the next section, the anisotropy will often evolve in time. Therefore, fine-tuning for a particular set of parameters is not necessarily useful in practice. Though we use the same $a,b$ for isotropic and anisotropic cases, we think it is sufficient because the dispersion relation obtained from this closure still predicts qualitatively the same behavior found in the fully kinetic model.

\section{Nonlinear simulation}
\label{sec4}
\subsection{Simulation model and numerical method}
In this section, we present nonlinear simulation results obtained by a model based on the proposed closure. Since the closure may independently be applied to any particle species, we consider only the ion dynamics for simplicity. In other words, electrons are treated as a massless, isothermal, and isotropic fluid. For ions, we solve Eqs. (\ref{continuity}-\ref{pressureeq}) with the non-local heat flux as detailed below. We assume that $p_{xx}=p_{yy}$ and $p_{xy}=0$ are always satisfied so that extra closure equations for these terms are not needed. We solve $p_{\perp}=(p_{xx}+p_{yy})/2$ instead of solving $p_{xx}$ and $p_{yy}$ separately. Henceforth, we denote $p_{zz}=p_{\parallel}$ for notation consistency. This makes the model an eight-moment model, while scalar and gyrotropic pressure (such as CGL-MHD) models may be categorized as five-moment and six-moment models, respectively.

The electric field is then determined by the generalized Ohm's law
\begin{equation}
\bm{E}+\bm{u}\times\bm{B}=-\frac{1}{en}\bm{B}\times(\nabla\times\bm{B})-\frac{T_e}{en}\nabla n
\end{equation}
where $T_e$ is the electron temperature and charge neutrality is assumed: $n=n_i\sim n_e$. We ignore the displacement current term of Eq. (\ref{M4}) to be consistent with the charge neurtality assumption. This provides a good approximation if we only consider wave length sufficiently longer than Debye length in the system whose electron plasma frequency is sufficiently larger than electron cyclotron frequency. The time evolution of the magnetic field is calculated by Eq. (\ref{M2}). If we use the standard adiabatic EoS for a scalar ion pressure, this model is identical to Hall-MHD.

We use a one-dimensional simulation box with the periodic boundary condition. All the physical quantities are transformed to Fourier coefficients using Fast Fourier Transform, then required derivatives and heat flux terms are calculated in the Fourier space; differentiation is simply multiplication by $ik$ in the Fourier space.

The transverse heat flux terms are calculated directly in Fourier space by Eq. (\ref{redefhf}) with
\begin{equation}
\begin{cases}
\nu=-iv_{\mathrm{th}}\frac{\sqrt{2\pi}}{\pi-2}\frac{k}{|k|},\\
\Pi=\frac{4-\pi}{\pi-2}.
\end{cases}
\end{equation}
The Landau closure is used for the longitudinal heat flux:
\begin{equation}
\tilde{q}_{zzz}=-i\sqrt{\frac{8}{\pi}}v_{\mathrm{th}}
\left(\tilde{p}_{\parallel}-\frac{p_{\parallel0}}{n_0}\tilde{n}\right)\frac{k}{|k|}.
\end{equation}
These quantities are brought back to $z$-space to evaluate the nonlinear terms via transform method. The resulting equations in the semi-discrete form are integrated by fourth-order explicit Runge-Kutta method.

\subsection{Simulation setup}
Here, we consider an instability driven by perpendicular temperature anisotropy $(T_{\perp}>T_{\parallel})$, which destabilizes EMIC waves. We use the same parameter as FIG. \ref{emic} $(\beta_{\perp}=1,\beta_{\parallel}=0.2)$. The initial condition is uniform in space except for small random noise added on to the transverse magnetic field ($\sim0.01\%$ of the ambient field). We use $\Delta z/\lambda_i=0.25, \Omega_{ci}\Delta t =0.01$ for space and time discretization. The size of the simulation box is $L_z=64\lambda_i$. The time integration is carried out up to $\Omega_{ci}t=800$.

The total energy of this system is the sum of ion pressure: $p_{\perp}+p_{\parallel}/2$, magnetic energy: $(B_x^2+B_y^2)/2\mu_0$, and kinetic energy: $nmu^2/2$. We have confirmed that the total energy is conserved within $0.1\%$ during the entire simulation interval even though it is not a conserved quantity in a strict sense because of the assumption of the isothermal EoS for the electron fluid.

\subsection{Results}
\begin{figure}[h]
\begin{center}
\includegraphics[width=0.75\linewidth]{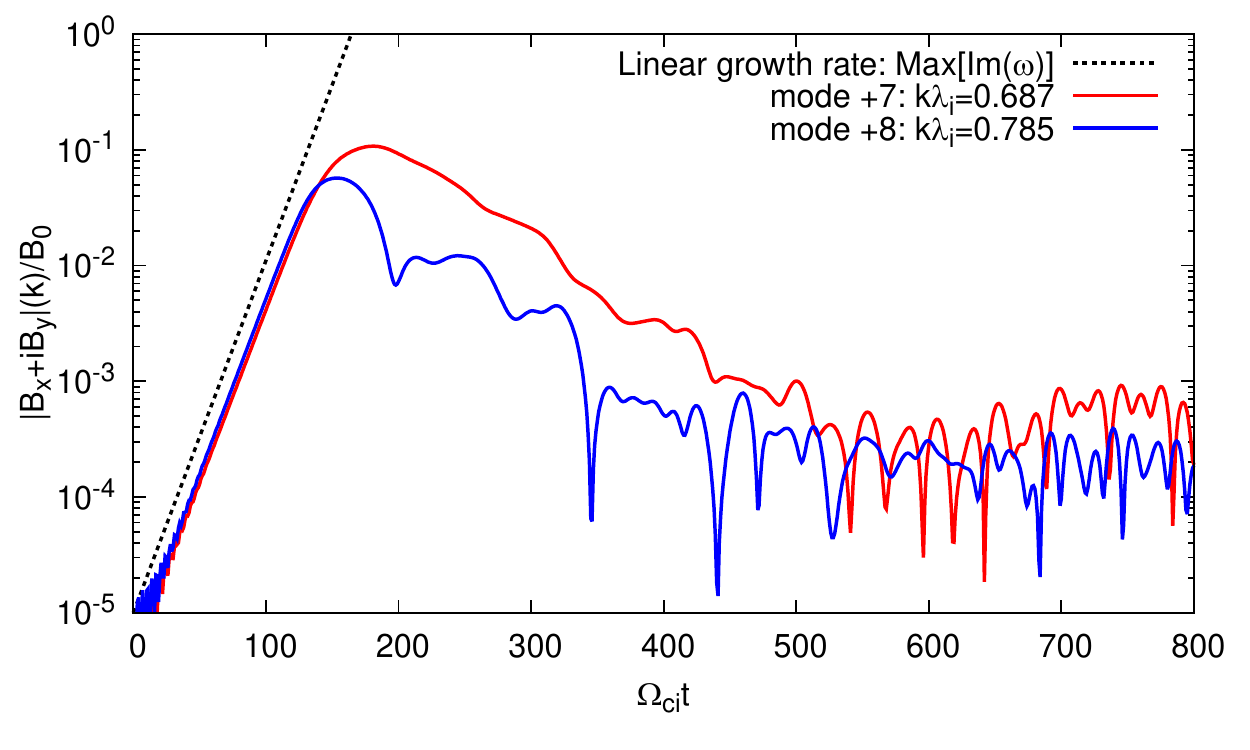}%
\caption{\label{growth} Comparison between simulation results and linear theory.
The time development of Fourier amplitude $|B_x+iB_y|(k)$ for mode $+7$ and $+8$ are shown.
Note that the theoretical growth rates for the two modes are nearly the same as the maximum growth rate.}
\end{center}
\end{figure}

\begin{figure}[h]
\begin{center}
\includegraphics[width=0.75\linewidth]{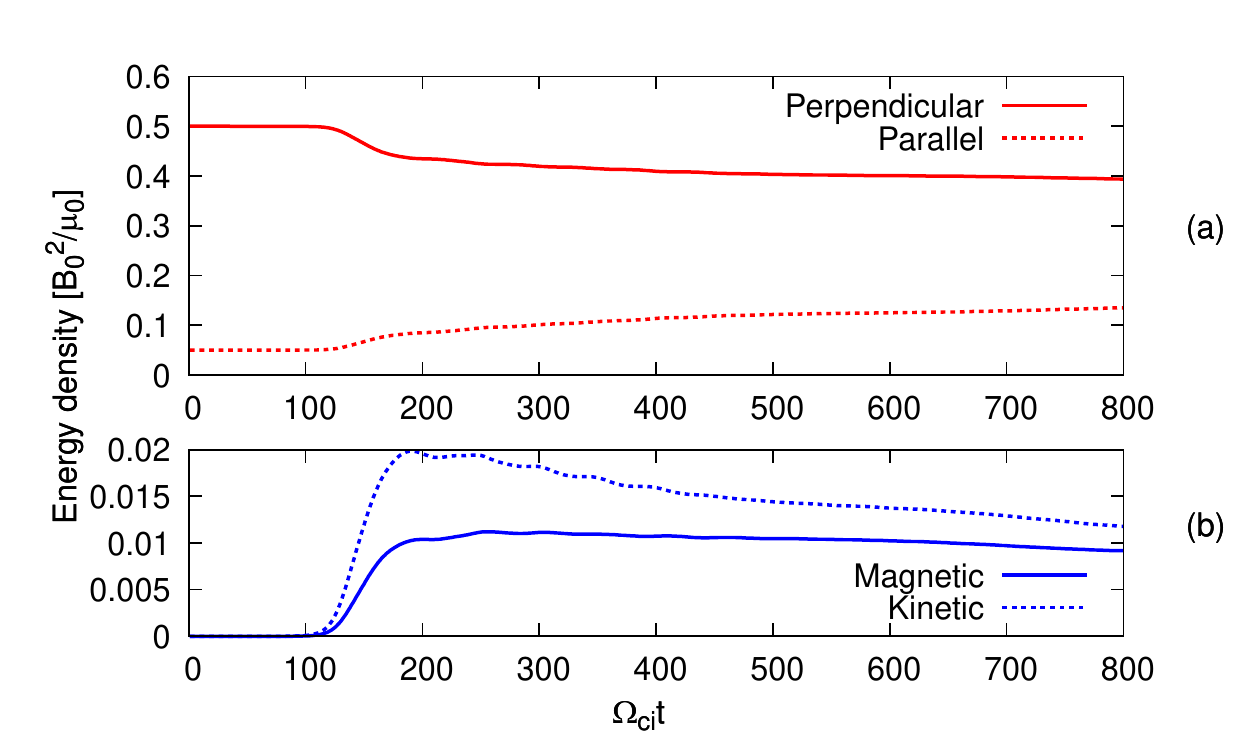}%
\caption{\label{energydensity} Evolution of energy density.
(a) Perpendicular $p_{\perp}$ and parallel $p_{\parallel}/2$ energy density.
(b) Transverse magnetic $(B_x^2+B_y^2)/2\mu_0$ (solid) and kinetic $nm(v_x^2+v_y^2+v_z^2)/2$ (dashed) energy density}
\end{center}
\end{figure}

\begin{figure}[h]
\begin{center}
\includegraphics[width=0.75\linewidth]{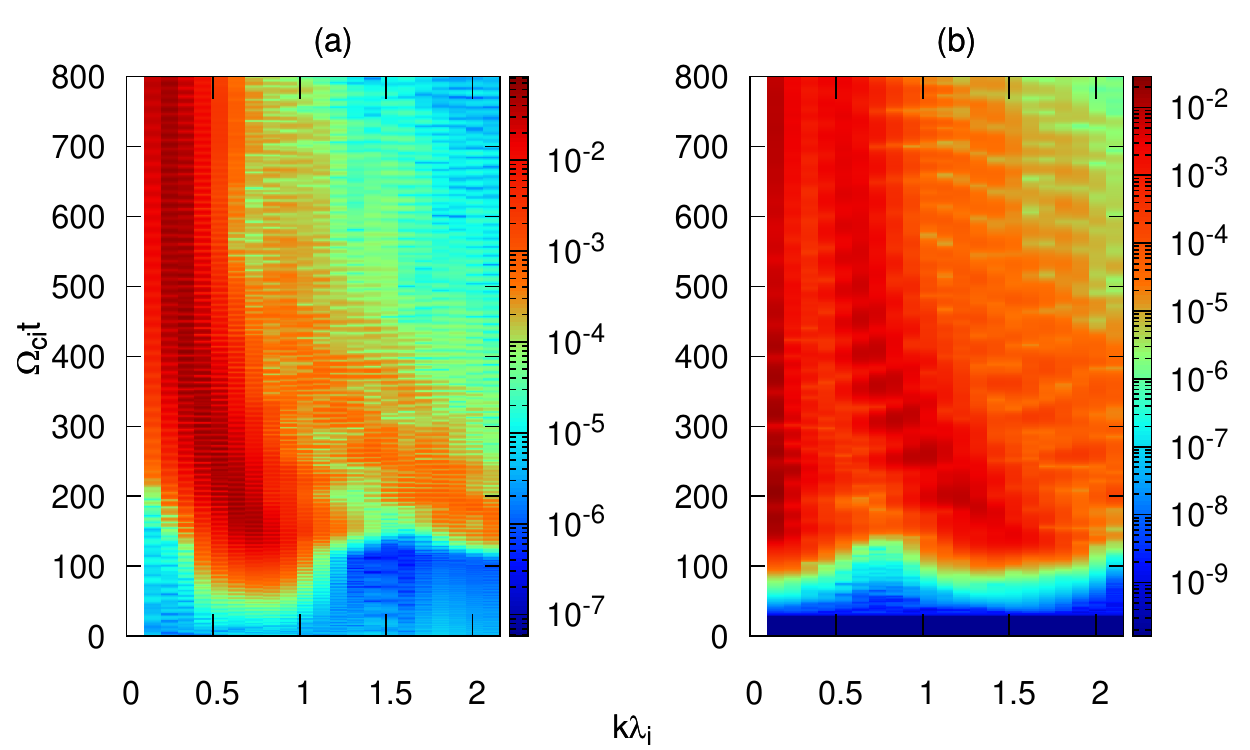}%
\caption{\label{spectre} Evolution of wavenumber spectrum.
(a) Magnetic field: $|\tilde{B}_x(k,t)|/B_0$. (b) Number density: $|\tilde{n}(k,t)|/n_0$.
The color is shown in logarithmic scale.}
\end{center}
\end{figure}

FIG. \ref{growth} shows that the simulation result is consistent with the linear theory. The initial development of modes with near the maximum growth rate ($+7$ and $+8$) matches almost exactly with the theoretical prediction. We can see that the linear approximation is valid during the initial phase $(0<\Omega_{ci}t<150)$. Although the closure is constructed for linearized equations and does not guarantee the nonlinear stability, the saturation of transverse magnetic field and relaxation of pressure anisotropy are observed in FIG. \ref{energydensity}, which might seem surprising at first glance. This qualitatively correct behavior of the saturation phase $(100<\Omega_{ci}t<200)$ will be discussed in the next subsection.

Long term evolution after the saturation can be seen in FIG. \ref{spectre}. FIG. \ref{spectre} (a) suggests that the magnetic energy is gradually transferred to longer wavelength modes. This inverse cascade can be explained as a result of successive operations of the parametric decay instability, or nonlinear wave-wave interactions \cite{Goldstein1978,Terasawa1986}. We see that such a nonlinear MHD/Hall-MHD process is successfully contained in the model. The coupling between transverse electromagnetic waves and acoustic waves can be seen in FIG. \ref{spectre} (b). Although the linear growth of acoustic waves is qualitatively similar between this model and Hall-MHD (see \citet{Nariyuki2006}, however, for the effect of linear Landau damping of acoustic waves), there is an important difference in nonlinear evolution. In a typical fluid where the dissipation occurs only at the grid scale, large-amplitude acoustic waves steepen to form shocks, and the heating of plasma takes place locally at the shock surfaces. We did not observe, however, such a signature of wave steepening in the density spectrum FIG. \ref{spectre} (b). In the present model, acoustic waves suffer Landau damping, which is taken into account approximately through the Landau closure. The dissipation thus happens globally without drastic steepening of waves. This is consistent with the characteristics of short wavelength acoustic waves in sufficiently high-$\beta$ plasmas.

\subsection{Quasilinear relaxation}
\label{QR}
Let us discuss how the isotropization observed in the simulation is explained. For this purpose, we consider the diagonal components of the pressure tensor equation and the equation for kinetic energy (which is obtained by taking the inner product between $mn\bm{u}$ and Eq. (\ref{EOM}) ).
\begin{align}
\frac{\partial p_{\perp}}{\partial t}+\frac{\partial}{\partial z}(p_{\perp}u_z)
+p_{xz}\frac{\partial u_x}{\partial z}+p_{yz}\frac{\partial u_y}{\partial z}
-\frac{e}{m}(B_xp_{yz}-B_yp_{xz}) &=0, \label{perp}\\
\frac{\partial p_{\parallel}}{\partial t}+\frac{\partial}{\partial z}(p_{\parallel}u_z+q_{zzz})
+2\left[p_{\parallel}\frac{\partial u_z}{\partial z}+\frac{e}{m}(B_xp_{yz}-B_yp_{xz})\right] &=0, \label{parallel}\\
\frac{\partial}{\partial t}\left(\frac{1}{2}mnu^2\right)
+\frac{\partial}{\partial z}\left(\frac{1}{2}mnu^2u_z+p_{xz}u_x+p_{yz}u_y+p_{\parallel}u_z\right)& \nonumber \\
-\left(p_{xz}\frac{\partial u_x}{\partial z}+p_{yz}\frac{\partial u_y}{\partial z}+p_{\parallel}\frac{\partial u_z}{\partial z}\right)
-ne\bm{u}\cdot\bm{E}&=0. \label{kinetic}
\end{align}
The second terms in these equations are in conservation form and they do not affect the spatially averaged pressures. $p_{zz}\frac{\partial u_z}{\partial z}$ is the term associated with the longitudinal mode and is not relevant for isotropization. The $p_{xz}\frac{\partial u_x}{\partial z}+p_{yz}\frac{\partial u_y}{\partial z}$ in Eq. (\ref{perp}) and Eq. (\ref{kinetic}) with the opposite signs, which may be understood as the energy exchange between perpendicular energy and kinetic energy via the instability.

The remaining term $B_xp_{yz}-B_yp_{xz}$ appears in both parallel and perpendicular equations with the opposite signs (recall that $2p_{\perp}=p_{xx}+p_{yy}$). Considering the total thermal energy is given by $\frac{1}{2}\mathrm{Tr}(\bm{p})=p_{\perp}+p_{\parallel}/2$, we see that this term effectively acts as the isotropization, which may reduce the initial anisotropy and suppress the wave growth in the nonlinear phase. Now let us investigate the nonlinear isotropization term in detail. The evolution of spatially averaged $(k=0)$ pressure is associated with:
\begin{equation}
\begin{split}
(B_xp_{yz}-B_yp_{xz})_{k=0}
&=\int dk' \left[\tilde{B}_x(k')\tilde{p}_{yz}(-k')
- \tilde{B}_y(k')\tilde{p}_{xz}(-k') \right]\\
&=-\int dk'\mathrm{Im}[\hat{B}(k')\hat{p}^{\ast}(k')]
\label{bcrossp}
\end{split}
\end{equation}
where $^\ast$ denotes complex conjugate. The relation between $\hat{B}$ and $\hat{p}$ is obtained from the linearized equations:
\begin{equation}
\hat{p}=\frac{e}{m}
\frac{p_{\parallel 0}\left[\omega\left(1+\frac{\Pi}{p_{\parallel 0}}\right)
+A(\omega-\Omega)\right]}
{(\omega-\Omega-k\nu)(\omega-\Omega)-k^2v_{\mathrm{th}}^2
\left(1+\frac{\Pi}{p_{\parallel 0}}\right)} \hat{B}.
\end{equation}
Note that the linear dispersion relation $\omega=\omega(k)$ must be used to evaluate the relation. Substituting this expression to Eq. (\ref{bcrossp}), we obtain the following quasilinear approximation for the isotropization term:
\begin{equation}
\begin{split}
&\frac{e}{m}(B_xp_{yz}-B_yp_{xz})_{k=0}
=\left(\frac{e}{m}\right)^2\int dk' |\hat{B}(k')|^2\\
&\times
\mathrm{Im}\left[\frac{p_{\parallel 0}\left[\omega\left(1+\frac{\Pi}{p_{\parallel 0}}\right)
+A(\omega-\Omega)\right]}
{(\omega-\Omega-k'\nu)(\omega-\Omega)-{k'}^2v_{\mathrm{th}}^2
\left(1+\frac{\Pi}{p_{\parallel 0}}\right)}\right].
\label{QLeval}
\end{split}
\end{equation}
We can see that this term is proportional to the wave magnetic energy. It is easy to confirm that the isotropization effect disappears if $\omega$ and $\nu$ were real valued. Our choice of the closure coefficients makes this term non-zero and results in the plausible isotropization. We should mention that these properties are very similar to the standard quasilinear theory for the fully kinetic model \citep{Yoon2017}.

\begin{figure}[H]
\begin{center}
\includegraphics[width=0.75\linewidth]{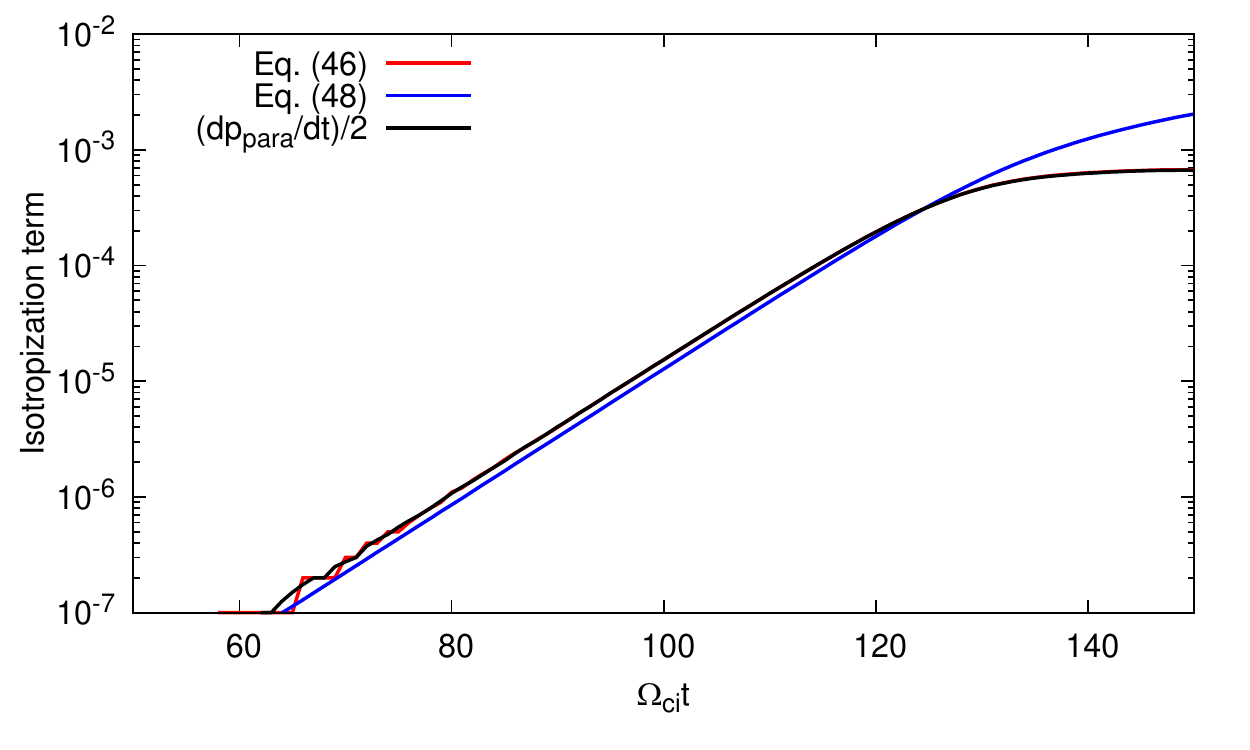}%
\caption{\label{qle} Comparison between isotropization term and temporal derivative of parallel pressure.
The direct evaluation of $-e/m(B_xp_{yz}-B_yp_{xz})_{k=0}$ is shown in red, while quasilinear approximation of Eq. (\ref{QLeval}) is in blue.
The black line indicating the temporal derivative $dp_{\parallel}/dt/2$ for comparison is nearly indistinguishable from the red line.}
\end{center}
\end{figure}

FIG.\ref{qle} compares the actual spatially averaged isotropization term of Eq. (\ref{bcrossp}) and the approximation Eq. (\ref{QLeval}) calculated by taking the sum of eight dominant modes seen in the simulation $(\pm 6,\pm 7, \pm 8, \pm 9)$. The coefficient was calculated using the initial condition. We can see that the change of parallel pressure (black) is indeed dominated by the isotropization term (red). The quasi-linear approximation shown in blue slightly underestimates the isotropization term, which comes from the contributions of the wavenumbers we did not include in the summation.

This comparison indicates the validity of the quasi-linear theory described here and also that the linear phase relationship between magnetic field $\hat{B}$ and the off-diagonal pressure $\hat{p}$ has been kept up to at around $\Omega_{ci}t=130$. At this time, the anisotropy is reduced substantially from the initial condition, which introduces non-negligible deviation to the linear eigenmodes of the system.

\section{Discussion and Conclusion}
\label{sec5}
We have discussed the method to take into account the cyclotron resonance effect in a fluid model. We have started from the moment equations derived from the Vlasov equation and determined the closure coefficients in wavenumber space so that the linear response for each particle species approximates the response obtained by linear kinetic theory assuming Maxwellian as the zeroth-order distribution function. We have successfully reproduced the qualitative behavior of dispersion relation described by $Z$-function.

Looking at this from a different perspective, we may understand that the model gives the exact response obtained for a different zeroth-order distribution which mimics Maxwellian distribution \cite{Hammett1990}. This raises the possibility that similar procedures extended to higher-order moments may give reasonable approximations to more complicated non-Maxwellian distributions. In principle, we may be able to incorporate the effect of nonlinear modification of the initial distribution function into the model if we find a scheme to systematically extend the closure model to higher-order terms, which will describe more complicated "shapes" of distribution functions.

The relaxation of the temperature anisotropy through the EMIC instability we saw in Sec.~\ref{sec4} can be regarded as one of the examples in which a non-local closure model succeeded in not only linear but also nonlinear regimes, reproducing a qualitatively correct long term behavior of an initially unstable system. The isotropization effect was fed back to the closure coefficients, which suppressed the wave growth and led to saturation. To the authors' knowledge, this kind of nonlinear feedback of kinetic instabilities has never been taken into account in a similar fluid model.

The original Landau closure \citep{Hammett1990} is capable of modeling the linear Landau damping of electrostatic waves for a Maxwellian plasma. It has also been extended to electromagnetic fluctuations in the low-frequency limit \citep{Snyder1997,Goswami2005,Passot2007,Passot2012,Sulem2015}. Although they differ from each other in detail, the non-local closure coefficients were determined solely based on linear Landau resonance for the gyrotropic components. The nonlinear effect associated with the deviation from the unperturbed distribution function does not seem to be included appropriately. Indeed, a typical strategy for nonlinear collisionless MHD simulations \citep{Sharma2006,Hirabayashi2016,Hirabayashi2017} is to include artificial pitch-angle scatterings, which will be turned on if the anisotropy develops into unstable regimes because otherwise, there is nothing to suppress the kinetic instability.

Our model, on the other hand, determines the closure componentwise for the response tensor; the existing Landau closure applied to the longitudinal component, and the new closure for transverse components. Since our model does not assume anything on the wave frequency, the non-gyrotropic components of the pressure tensor cannot be ignored. We actually find that they play the essential role for the cyclotron resonance effect. The closure model proposed here gives a reasonable approximation up to the cyclotron frequency of the resonant particle species. It is a rather unexpected finding that the non-gyrotropic components, when appropriately modeled with a non-local closure, contribute to the relaxation of anisotropy. We think this result is encouraging because our approach is quite straightforward and may be extended to higher orders.

It is worth mentioning that a Landau-type closure has also been applied to the collisionless magnetic reconnection problem. It is well known that the electron non-gyrotropy is essential for breaking the magnetic field frozen-in condition in the electron diffusion region. \citet{Wang2015} and subsequent works \citep{Ng2017,Ng2020} used a ten-moment model with the full pressure tensor in their two-fluid (ions and electrons) simulations of magnetic reconnection. They naively applied the original electrostatic Landau closure for the scalar pressure to the full $3 \times 3$ tensor, which, however, does not have a solid theoretical basis. The philosophy of Landau-type closure is to approximate the fully kinetic heat flux by a linear combination of lower-order moments. The componentwise approach taken in this paper directly follows this. For the problem of collisionless magnetic reconnection, one of the possibilities is to approximate the heat flux to mimic the electron response in the linear collisionless tearing mode where the electron inertial resistivity plays the essential role.

Finally, let us discuss the potential of non-local closure fluid models for macroscopic simulations. To apply a non-local closure model to a general system that requires beyond-MHD descriptions but is too demanding for particle codes, further extensions of the model will be necessary. Obviously, an extension to higher-order moments is useful to represent more complicated distribution functions. In addition, proper treatment of oblique propagation and inhomogeneous equilibrium states needs to be worked out. The former can presumably be achieved by adopting some kind of finite-Larmor-radius (FLR) correction. Landau-fluid-based models with FLR corrections \cite{Goswami2005,Sulem2015,Sarto_2017} have a certain success in achieving a better agreement with the kinetic result for obliquely propagating modes such as the mirror instability.
Details of FLR models and application to parallel and oblique firehose instabilities are given in \citep{Hunana2019}.
The latter may require further theoretical consideration and better numerical methods.

\begin{acknowledgments}
This work was supported by JSPS KAKENHI grant Nos.~17H02966 and 17H06140.
T. J. was supported by International Graduate Program for Excellence in Earth-Space Science (IGPEES),
The University of Tokyo.
\end{acknowledgments}

\section*{data availability}
The data that support the findings of this study are available from the corresponding author
upon reasonable request.

\nocite{*}

\begin{thebibliography}{26}%
\makeatletter
\providecommand \@ifxundefined [1]{%
 \@ifx{#1\undefined}
}%
\providecommand \@ifnum [1]{%
 \ifnum #1\expandafter \@firstoftwo
 \else \expandafter \@secondoftwo
 \fi
}%
\providecommand \@ifx [1]{%
 \ifx #1\expandafter \@firstoftwo
 \else \expandafter \@secondoftwo
 \fi
}%
\providecommand \natexlab [1]{#1}%
\providecommand \enquote  [1]{``#1''}%
\providecommand \bibnamefont  [1]{#1}%
\providecommand \bibfnamefont [1]{#1}%
\providecommand \citenamefont [1]{#1}%
\providecommand \href@noop [0]{\@secondoftwo}%
\providecommand \href [0]{\begingroup \@sanitize@url \@href}%
\providecommand \@href[1]{\@@startlink{#1}\@@href}%
\providecommand \@@href[1]{\endgroup#1\@@endlink}%
\providecommand \@sanitize@url [0]{\catcode `\\12\catcode `\$12\catcode
  `\&12\catcode `\#12\catcode `\^12\catcode `\_12\catcode `\%12\relax}%
\providecommand \@@startlink[1]{}%
\providecommand \@@endlink[0]{}%
\providecommand \url  [0]{\begingroup\@sanitize@url \@url }%
\providecommand \@url [1]{\endgroup\@href {#1}{\urlprefix }}%
\providecommand \urlprefix  [0]{URL }%
\providecommand \Eprint [0]{\href }%
\providecommand \doibase [0]{http://dx.doi.org/}%
\providecommand \selectlanguage [0]{\@gobble}%
\providecommand \bibinfo  [0]{\@secondoftwo}%
\providecommand \bibfield  [0]{\@secondoftwo}%
\providecommand \translation [1]{[#1]}%
\providecommand \BibitemOpen [0]{}%
\providecommand \bibitemStop [0]{}%
\providecommand \bibitemNoStop [0]{.\EOS\space}%
\providecommand \EOS [0]{\spacefactor3000\relax}%
\providecommand \BibitemShut  [1]{\csname bibitem#1\endcsname}%
\let\auto@bib@innerbib\@empty
\bibitem [{\citenamefont {Hammett}\ and\ \citenamefont
  {Perkins}(1990)}]{Hammett1990}%
  \BibitemOpen
  \bibfield  {author} {\bibinfo {author} {\bibfnamefont {G.~W.}\ \bibnamefont
  {Hammett}}\ and\ \bibinfo {author} {\bibfnamefont {F.~W.}\ \bibnamefont
  {Perkins}},\ }\bibfield  {title} {\enquote {\bibinfo {title} {Fluid moment
  models for landau damping with application to the ion-temperature-gradient
  instability},}\ }\href {\doibase 10.1103/PhysRevLett.64.3019} {\bibfield
  {journal} {\bibinfo  {journal} {Phys. Rev. Lett.}\ }\textbf {\bibinfo
  {volume} {64}},\ \bibinfo {pages} {3019--3022} (\bibinfo {year}
  {1990})}\BibitemShut {NoStop}%
\bibitem [{\citenamefont {Hammett}, \citenamefont {Dorland},\ and\
  \citenamefont {Perkins}(1992)}]{Hammett1992}%
  \BibitemOpen
  \bibfield  {author} {\bibinfo {author} {\bibfnamefont {G.~W.}\ \bibnamefont
  {Hammett}}, \bibinfo {author} {\bibfnamefont {W.}~\bibnamefont {Dorland}}, \
  and\ \bibinfo {author} {\bibfnamefont {F.~W.}\ \bibnamefont {Perkins}},\
  }\bibfield  {title} {\enquote {\bibinfo {title} {Fluid models of phase
  mixing, landau damping, and nonlinear gyrokinetic dynamics},}\ }\href
  {\doibase 10.1063/1.860014} {\bibfield  {journal} {\bibinfo  {journal}
  {Physics of Fluids B: Plasma Physics}\ }\textbf {\bibinfo {volume} {4}},\
  \bibinfo {pages} {2052--2061} (\bibinfo {year} {1992})}\BibitemShut {NoStop}%
\bibitem [{\citenamefont {Snyder}, \citenamefont {Hammett},\ and\ \citenamefont
  {Dorland}(1997)}]{Snyder1997}%
  \BibitemOpen
  \bibfield  {author} {\bibinfo {author} {\bibfnamefont {P.~B.}\ \bibnamefont
  {Snyder}}, \bibinfo {author} {\bibfnamefont {G.~W.}\ \bibnamefont {Hammett}},
  \ and\ \bibinfo {author} {\bibfnamefont {W.}~\bibnamefont {Dorland}},\
  }\bibfield  {title} {\enquote {\bibinfo {title} {Landau fluid models of
  collisionless magnetohydrodynamics},}\ }\href {\doibase 10.1063/1.872517}
  {\bibfield  {journal} {\bibinfo  {journal} {Physics of Plasmas}\ }\textbf
  {\bibinfo {volume} {4}},\ \bibinfo {pages} {3974--3985} (\bibinfo {year}
  {1997})}\BibitemShut {NoStop}%
\bibitem [{\citenamefont {Chew}, \citenamefont {Goldberger},\ and\
  \citenamefont {Low}(1956)}]{Chew1956}%
  \BibitemOpen
  \bibfield  {author} {\bibinfo {author} {\bibfnamefont {G.~F.}\ \bibnamefont
  {Chew}}, \bibinfo {author} {\bibfnamefont {M.~L.}\ \bibnamefont
  {Goldberger}}, \ and\ \bibinfo {author} {\bibfnamefont {F.~E.}\ \bibnamefont
  {Low}},\ }\bibfield  {title} {\enquote {\bibinfo {title} {The boltzmann
  equation and the one-fluid hydromagnetic equations in the absence of particle
  collisions},}\ }\href {\doibase 10.1098/rspa.1956.0116} {\bibfield  {journal}
  {\bibinfo  {journal} {Proceedings of the Royal Society A: Mathematical,
  Physical and Engineering Sciences}\ }\textbf {\bibinfo {volume} {236}},\
  \bibinfo {pages} {112--118} (\bibinfo {year} {1956})}\BibitemShut {NoStop}%
\bibitem [{\citenamefont {Sharma}, \citenamefont {Hammett},\ and\ \citenamefont
  {Quataert}(2003)}]{Sharma2003}%
  \BibitemOpen
  \bibfield  {author} {\bibinfo {author} {\bibfnamefont {P.}~\bibnamefont
  {Sharma}}, \bibinfo {author} {\bibfnamefont {G.~W.}\ \bibnamefont {Hammett}},
  \ and\ \bibinfo {author} {\bibfnamefont {E.}~\bibnamefont {Quataert}},\
  }\bibfield  {title} {\enquote {\bibinfo {title} {Transition from
  collisionless to collisional magnetorotational instability},}\ }\href
  {\doibase 10.1086/378234} {\bibfield  {journal} {\bibinfo  {journal} {The
  Astrophysical Journal}\ }\textbf {\bibinfo {volume} {596}} (\bibinfo {year}
  {2003}),\ 10.1086/378234}\BibitemShut {NoStop}%
\bibitem [{\citenamefont {Sharma}\ \emph {et~al.}(2006)\citenamefont {Sharma},
  \citenamefont {Hammett}, \citenamefont {Quataert},\ and\ \citenamefont
  {Stone}}]{Sharma2006}%
  \BibitemOpen
  \bibfield  {author} {\bibinfo {author} {\bibfnamefont {P.}~\bibnamefont
  {Sharma}}, \bibinfo {author} {\bibfnamefont {G.~W.}\ \bibnamefont {Hammett}},
  \bibinfo {author} {\bibfnamefont {E.}~\bibnamefont {Quataert}}, \ and\
  \bibinfo {author} {\bibfnamefont {J.~M.}\ \bibnamefont {Stone}},\ }\bibfield
  {title} {\enquote {\bibinfo {title} {Shearing box simulations of the mri in a
  collisionless plasma},}\ }\href {\doibase 10.1086/498405} {\bibfield
  {journal} {\bibinfo  {journal} {The Astrophysical Journal}\ }\textbf
  {\bibinfo {volume} {637}},\ \bibinfo {pages} {952--967} (\bibinfo {year}
  {2006})}\BibitemShut {NoStop}%
\bibitem [{\citenamefont {Goswami}, \citenamefont {Passot},\ and\ \citenamefont
  {Sulem}(2005)}]{Goswami2005}%
  \BibitemOpen
  \bibfield  {author} {\bibinfo {author} {\bibfnamefont {P.}~\bibnamefont
  {Goswami}}, \bibinfo {author} {\bibfnamefont {T.}~\bibnamefont {Passot}}, \
  and\ \bibinfo {author} {\bibfnamefont {P.~L.}\ \bibnamefont {Sulem}},\
  }\bibfield  {title} {\enquote {\bibinfo {title} {A landau fluid model for
  warm collisionless plasmas},}\ }\href {\doibase 10.1063/1.2096582} {\bibfield
   {journal} {\bibinfo  {journal} {Physics of Plasmas}\ }\textbf {\bibinfo
  {volume} {12}} (\bibinfo {year} {2005}),\ 10.1063/1.2096582}\BibitemShut
  {NoStop}%
\bibitem [{\citenamefont {Passot}\ and\ \citenamefont
  {Sulem}(2007)}]{Passot2007}%
  \BibitemOpen
  \bibfield  {author} {\bibinfo {author} {\bibfnamefont {T.}~\bibnamefont
  {Passot}}\ and\ \bibinfo {author} {\bibfnamefont {P.~L.}\ \bibnamefont
  {Sulem}},\ }\bibfield  {title} {\enquote {\bibinfo {title} {Collisionless
  magnetohydrodynamics with gyrokinetic effects},}\ }\href {\doibase
  10.1063/1.2751601} {\bibfield  {journal} {\bibinfo  {journal} {Physics of
  Plasmas}\ }\textbf {\bibinfo {volume} {14}},\ \bibinfo {pages} {082502}
  (\bibinfo {year} {2007})}\BibitemShut {NoStop}%
\bibitem [{\citenamefont {Passot}, \citenamefont {Sulem},\ and\ \citenamefont
  {Hunana}(2012)}]{Passot2012}%
  \BibitemOpen
  \bibfield  {author} {\bibinfo {author} {\bibfnamefont {T.}~\bibnamefont
  {Passot}}, \bibinfo {author} {\bibfnamefont {P.~L.}\ \bibnamefont {Sulem}}, \
  and\ \bibinfo {author} {\bibfnamefont {P.}~\bibnamefont {Hunana}},\
  }\bibfield  {title} {\enquote {\bibinfo {title} {Extending
  magnetohydrodynamics to the slow dynamics of collisionless plasmas},}\ }\href
  {\doibase 10.1063/1.4746092} {\bibfield  {journal} {\bibinfo  {journal}
  {Physics of Plasmas}\ }\textbf {\bibinfo {volume} {19}},\ \bibinfo {pages}
  {082113} (\bibinfo {year} {2012})}\BibitemShut {NoStop}%
\bibitem [{\citenamefont {Sulem}\ and\ \citenamefont
  {Passot}(2015)}]{Sulem2015}%
  \BibitemOpen
  \bibfield  {author} {\bibinfo {author} {\bibfnamefont {P.~L.}\ \bibnamefont
  {Sulem}}\ and\ \bibinfo {author} {\bibfnamefont {T.}~\bibnamefont {Passot}},\
  }\bibfield  {title} {\enquote {\bibinfo {title} {Landau fluid closures with
  nonlinear large-scale finite larmor radius corrections for collisionless
  plasmas},}\ }\href {\doibase 10.1017/S0022377814000671} {\bibfield  {journal}
  {\bibinfo  {journal} {Journal of Plasma Physics}\ }\textbf {\bibinfo {volume}
  {81}},\ \bibinfo {pages} {325810103} (\bibinfo {year} {2015})}\BibitemShut
  {NoStop}%
\bibitem [{\citenamefont {Hesse}\ and\ \citenamefont
  {Winske}(1993)}]{Hesse1993}%
  \BibitemOpen
  \bibfield  {author} {\bibinfo {author} {\bibfnamefont {M.}~\bibnamefont
  {Hesse}}\ and\ \bibinfo {author} {\bibfnamefont {D.}~\bibnamefont {Winske}},\
  }\bibfield  {title} {\enquote {\bibinfo {title} {Hybrid simulations of
  collisionless ion tearing},}\ }\href {\doibase 10.1029/93GL01250} {\bibfield
  {journal} {\bibinfo  {journal} {Geophysical Research Letters}\ }\textbf
  {\bibinfo {volume} {20}},\ \bibinfo {pages} {1207--1210} (\bibinfo {year}
  {1993})}\BibitemShut {NoStop}%
\bibitem [{\citenamefont {Hesse}\ and\ \citenamefont
  {Winske}(1994)}]{Hesse1994}%
  \BibitemOpen
  \bibfield  {author} {\bibinfo {author} {\bibfnamefont {M.}~\bibnamefont
  {Hesse}}\ and\ \bibinfo {author} {\bibfnamefont {D.}~\bibnamefont {Winske}},\
  }\bibfield  {title} {\enquote {\bibinfo {title} {Hybrid simulations of
  collisionless reconnection in current sheets},}\ }\href {\doibase
  10.1029/94JA00676} {\bibfield  {journal} {\bibinfo  {journal} {Journal of
  Geophysical Research}\ }\textbf {\bibinfo {volume} {99}},\ \bibinfo {pages}
  {11177} (\bibinfo {year} {1994})}\BibitemShut {NoStop}%
\bibitem [{\citenamefont {{Stix}}(1992)}]{Stix}%
  \BibitemOpen
  \bibfield  {author} {\bibinfo {author} {\bibfnamefont {T.~H.}\ \bibnamefont
  {{Stix}}},\ }\href@noop {} {\emph {\bibinfo {title} {{Waves in plasmas}}}}\
  (\bibinfo {year} {1992})\BibitemShut {NoStop}%
\bibitem [{\citenamefont {Davidson}\ and\ \citenamefont
  {Ogden}(1975)}]{Davidson1975}%
  \BibitemOpen
  \bibfield  {author} {\bibinfo {author} {\bibfnamefont {R.~C.}\ \bibnamefont
  {Davidson}}\ and\ \bibinfo {author} {\bibfnamefont {J.~M.}\ \bibnamefont
  {Ogden}},\ }\bibfield  {title} {\enquote {\bibinfo {title} {Electromagnetic
  ion cyclotron instability driven by ion energy anisotropy in high‐beta
  plasmas},}\ }\href {\doibase 10.1063/1.861253} {\bibfield  {journal}
  {\bibinfo  {journal} {The Physics of Fluids}\ }\textbf {\bibinfo {volume}
  {18}},\ \bibinfo {pages} {1045--1050} (\bibinfo {year} {1975})}\BibitemShut
  {NoStop}%
\bibitem [{\citenamefont {{Goldstein}}(1978)}]{Goldstein1978}%
  \BibitemOpen
  \bibfield  {author} {\bibinfo {author} {\bibfnamefont {M.~L.}\ \bibnamefont
  {{Goldstein}}},\ }\bibfield  {title} {\enquote {\bibinfo {title} {{An
  instability of finite amplitude circularly polarized Afv{\'e}n waves.}}}\
  }\href {\doibase 10.1086/155829} {\bibfield  {journal} {\bibinfo  {journal}
  {\apj}\ }\textbf {\bibinfo {volume} {219}},\ \bibinfo {pages} {700--704}
  (\bibinfo {year} {1978})}\BibitemShut {NoStop}%
\bibitem [{\citenamefont {Terasawa}\ \emph {et~al.}(1986)\citenamefont
  {Terasawa}, \citenamefont {Hoshino}, \citenamefont {Sakai},\ and\
  \citenamefont {Hada}}]{Terasawa1986}%
  \BibitemOpen
  \bibfield  {author} {\bibinfo {author} {\bibfnamefont {T.}~\bibnamefont
  {Terasawa}}, \bibinfo {author} {\bibfnamefont {M.}~\bibnamefont {Hoshino}},
  \bibinfo {author} {\bibfnamefont {J.-I.}\ \bibnamefont {Sakai}}, \ and\
  \bibinfo {author} {\bibfnamefont {T.}~\bibnamefont {Hada}},\ }\bibfield
  {title} {\enquote {\bibinfo {title} {Decay instability of finite-amplitude
  circularly polarized alfven waves - a numerical simulation of stimulated
  brillouin scattering},}\ }\href@noop {} {\bibfield  {journal} {\bibinfo
  {journal} {Journal of Geophysical Research}\ }\textbf {\bibinfo {volume}
  {91}},\ \bibinfo {pages} {4171--4187} (\bibinfo {year} {1986})}\BibitemShut
  {NoStop}%
\bibitem [{\citenamefont {Nariyuki}\ and\ \citenamefont
  {Hada}(2006)}]{Nariyuki2006}%
  \BibitemOpen
  \bibfield  {author} {\bibinfo {author} {\bibfnamefont {Y.}~\bibnamefont
  {Nariyuki}}\ and\ \bibinfo {author} {\bibfnamefont {T.}~\bibnamefont
  {Hada}},\ }\bibfield  {title} {\enquote {\bibinfo {title} {Kinetically
  modified parametric instabilities of circularly polarized alfvén waves: Ion
  kinetic effects},}\ }\href {\doibase 10.1063/1.2399468} {\bibfield  {journal}
  {\bibinfo  {journal} {Physics of Plasmas}\ }\textbf {\bibinfo {volume} {13}}
  (\bibinfo {year} {2006}),\ 10.1063/1.2399468}\BibitemShut {NoStop}%
\bibitem [{\citenamefont {Yoon}(2017)}]{Yoon2017}%
  \BibitemOpen
  \bibfield  {author} {\bibinfo {author} {\bibfnamefont {P.~H.}\ \bibnamefont
  {Yoon}},\ }\bibfield  {title} {\enquote {\bibinfo {title} {Kinetic
  instabilities in the solar wind driven by temperature anisotropies},}\ }\href
  {\doibase 10.1007/s41614-017-0006-1} {\bibfield  {journal} {\bibinfo
  {journal} {Reviews of Modern Plasma Physics}\ }\textbf {\bibinfo {volume}
  {1}},\ \bibinfo {pages} {4} (\bibinfo {year} {2017})}\BibitemShut {NoStop}%
\bibitem [{\citenamefont {Hirabayashi}, \citenamefont {Hoshino},\ and\
  \citenamefont {Amano}(2016)}]{Hirabayashi2016}%
  \BibitemOpen
  \bibfield  {author} {\bibinfo {author} {\bibfnamefont {K.}~\bibnamefont
  {Hirabayashi}}, \bibinfo {author} {\bibfnamefont {M.}~\bibnamefont
  {Hoshino}}, \ and\ \bibinfo {author} {\bibfnamefont {T.}~\bibnamefont
  {Amano}},\ }\bibfield  {title} {\enquote {\bibinfo {title} {A new framework
  for magnetohydrodynamic simulations with anisotropic pressure},}\ }\href
  {\doibase 10.1016/j.jcp.2016.09.064} {\bibfield  {journal} {\bibinfo
  {journal} {Journal of Computational Physics}\ }\textbf {\bibinfo {volume}
  {327}},\ \bibinfo {pages} {851--872} (\bibinfo {year} {2016})}\BibitemShut
  {NoStop}%
\bibitem [{\citenamefont {Hirabayashi}\ and\ \citenamefont
  {Hoshino}(2017)}]{Hirabayashi2017}%
  \BibitemOpen
  \bibfield  {author} {\bibinfo {author} {\bibfnamefont {K.}~\bibnamefont
  {Hirabayashi}}\ and\ \bibinfo {author} {\bibfnamefont {M.}~\bibnamefont
  {Hoshino}},\ }\bibfield  {title} {\enquote {\bibinfo {title} {Stratified
  simulations of collisionless accretion disks},}\ }\href {\doibase
  10.3847/1538-4357/aa74b3} {\bibfield  {journal} {\bibinfo  {journal} {The
  Astrophysical Journal}\ }\textbf {\bibinfo {volume} {842}},\ \bibinfo {pages}
  {36} (\bibinfo {year} {2017})}\BibitemShut {NoStop}%
\bibitem [{\citenamefont {Wang}\ \emph {et~al.}(2015)\citenamefont {Wang},
  \citenamefont {Hakim}, \citenamefont {Bhattacharjee},\ and\ \citenamefont
  {Germaschewski}}]{Wang2015}%
  \BibitemOpen
  \bibfield  {author} {\bibinfo {author} {\bibfnamefont {L.}~\bibnamefont
  {Wang}}, \bibinfo {author} {\bibfnamefont {A.~H.}\ \bibnamefont {Hakim}},
  \bibinfo {author} {\bibfnamefont {A.}~\bibnamefont {Bhattacharjee}}, \ and\
  \bibinfo {author} {\bibfnamefont {K.}~\bibnamefont {Germaschewski}},\
  }\bibfield  {title} {\enquote {\bibinfo {title} {Comparison of multi-fluid
  moment models with particle-in-cell simulations of collisionless magnetic
  reconnection},}\ }\href {\doibase 10.1063/1.4906063} {\bibfield  {journal}
  {\bibinfo  {journal} {Physics of Plasmas}\ }\textbf {\bibinfo {volume}
  {22}},\ \bibinfo {pages} {012108} (\bibinfo {year} {2015})}\BibitemShut
  {NoStop}%
\bibitem [{\citenamefont {Ng}\ \emph {et~al.}(2017)\citenamefont {Ng},
  \citenamefont {Hakim}, \citenamefont {Bhattacharjee}, \citenamefont
  {Stanier},\ and\ \citenamefont {Daughton}}]{Ng2017}%
  \BibitemOpen
  \bibfield  {author} {\bibinfo {author} {\bibfnamefont {J.}~\bibnamefont
  {Ng}}, \bibinfo {author} {\bibfnamefont {A.}~\bibnamefont {Hakim}}, \bibinfo
  {author} {\bibfnamefont {A.}~\bibnamefont {Bhattacharjee}}, \bibinfo {author}
  {\bibfnamefont {A.}~\bibnamefont {Stanier}}, \ and\ \bibinfo {author}
  {\bibfnamefont {W.}~\bibnamefont {Daughton}},\ }\bibfield  {title} {\enquote
  {\bibinfo {title} {Simulations of anti-parallel reconnection using a nonlocal
  heat flux closure},}\ }\href {\doibase 10.1063/1.4993195} {\bibfield
  {journal} {\bibinfo  {journal} {Physics of Plasmas}\ }\textbf {\bibinfo
  {volume} {24}},\ \bibinfo {pages} {082112} (\bibinfo {year}
  {2017})}\BibitemShut {NoStop}%
\bibitem [{\citenamefont {Ng}\ \emph {et~al.}(2020)\citenamefont {Ng},
  \citenamefont {Hakim}, \citenamefont {Wang},\ and\ \citenamefont
  {Bhattacharjee}}]{Ng2020}%
  \BibitemOpen
  \bibfield  {author} {\bibinfo {author} {\bibfnamefont {J.}~\bibnamefont
  {Ng}}, \bibinfo {author} {\bibfnamefont {A.}~\bibnamefont {Hakim}}, \bibinfo
  {author} {\bibfnamefont {L.}~\bibnamefont {Wang}}, \ and\ \bibinfo {author}
  {\bibfnamefont {A.}~\bibnamefont {Bhattacharjee}},\ }\bibfield  {title}
  {\enquote {\bibinfo {title} {An improved ten-moment closure for reconnection
  and instabilities},}\ }\href {\doibase 10.1063/5.0012067} {\bibfield
  {journal} {\bibinfo  {journal} {Physics of Plasmas}\ }\textbf {\bibinfo
  {volume} {27}} (\bibinfo {year} {2020}),\ 10.1063/5.0012067}\BibitemShut
  {NoStop}%
\bibitem [{\citenamefont {Sarto}, \citenamefont {Pegoraro},\ and\ \citenamefont
  {Tenerani}(2017)}]{Sarto_2017}%
  \BibitemOpen
  \bibfield  {author} {\bibinfo {author} {\bibfnamefont {D.~D.}\ \bibnamefont
  {Sarto}}, \bibinfo {author} {\bibfnamefont {F.}~\bibnamefont {Pegoraro}}, \
  and\ \bibinfo {author} {\bibfnamefont {A.}~\bibnamefont {Tenerani}},\
  }\bibfield  {title} {\enquote {\bibinfo {title} {`magneto-elastic' waves in
  an anisotropic magnetised plasma},}\ }\href {\doibase
  10.1088/1361-6587/aa56bd} {\bibfield  {journal} {\bibinfo  {journal} {Plasma
  Physics and Controlled Fusion}\ }\textbf {\bibinfo {volume} {59}},\ \bibinfo
  {pages} {045002} (\bibinfo {year} {2017})}\BibitemShut {NoStop}%
\bibitem [{\citenamefont {Hunana}\ \emph {et~al.}(2019)\citenamefont {Hunana},
  \citenamefont {Tenerani}, \citenamefont {Zank}, \citenamefont {Khomenko},
  \citenamefont {Goldstein}, \citenamefont {Webb}, \citenamefont {Cally},
  \citenamefont {Collados}, \citenamefont {Velli}, \citenamefont {Adhikari},\
  and\ \citenamefont {et~al.}}]{Hunana2019}%
  \BibitemOpen
  \bibfield  {author} {\bibinfo {author} {\bibfnamefont {P.}~\bibnamefont
  {Hunana}}, \bibinfo {author} {\bibfnamefont {A.}~\bibnamefont {Tenerani}},
  \bibinfo {author} {\bibfnamefont {G.~P.}\ \bibnamefont {Zank}}, \bibinfo
  {author} {\bibfnamefont {E.}~\bibnamefont {Khomenko}}, \bibinfo {author}
  {\bibfnamefont {M.~L.}\ \bibnamefont {Goldstein}}, \bibinfo {author}
  {\bibfnamefont {G.~M.}\ \bibnamefont {Webb}}, \bibinfo {author}
  {\bibfnamefont {P.~S.}\ \bibnamefont {Cally}}, \bibinfo {author}
  {\bibfnamefont {M.}~\bibnamefont {Collados}}, \bibinfo {author}
  {\bibfnamefont {M.}~\bibnamefont {Velli}}, \bibinfo {author} {\bibfnamefont
  {L.}~\bibnamefont {Adhikari}}, \ and\ \bibinfo {author} {\bibnamefont
  {et~al.}},\ }\bibfield  {title} {\enquote {\bibinfo {title} {An introductory
  guide to fluid models with anisotropic temperatures. part 1. cgl description
  and collisionless fluid hierarchy},}\ }\href {\doibase
  10.1017/S0022377819000801} {\bibfield  {journal} {\bibinfo  {journal}
  {Journal of Plasma Physics}\ }\textbf {\bibinfo {volume} {85}},\ \bibinfo
  {pages} {205850602} (\bibinfo {year} {2019})}\BibitemShut {NoStop}%
\bibitem [{\citenamefont {Chust}\ and\ \citenamefont
  {Belmont}(2006)}]{Chust2006}%
  \BibitemOpen
  \bibfield  {author} {\bibinfo {author} {\bibfnamefont {T.}~\bibnamefont
  {Chust}}\ and\ \bibinfo {author} {\bibfnamefont {G.}~\bibnamefont
  {Belmont}},\ }\bibfield  {title} {\enquote {\bibinfo {title} {Closure of
  fluid equations in collisionless magnetoplasmas},}\ }\href {\doibase
  10.1063/1.2138568} {\bibfield  {journal} {\bibinfo  {journal} {Physics of
  Plasmas}\ }\textbf {\bibinfo {volume} {13}},\ \bibinfo {pages} {012506}
  (\bibinfo {year} {2006})}\BibitemShut {NoStop}%
\end{thebibliography}
\providecommand{\noopsort}[1]{}\providecommand{\singleletter}[1]{#1}%

\end{document}